\documentclass{article}

\usepackage{hyperref}
\usepackage{color}
\usepackage{graphicx}
\usepackage{amstext,amsmath,amssymb,amsfonts,bbm, amsthm}
\usepackage{fancyhdr}

\begin{document}

\title{Aspects of the BTZ black hole interacting with fields}

\author{Leonardo Ort\'{i}z\footnote{lortiz@fis.cinvestav.mx} and 
Nora Bret\'{o}n\footnote{nora@fis.cinvestav.mx} \vspace{1cm}\\
Departamento de F\'{i}sica \\
Centro de Investigaci\'{o}n y de Estudios Avanzados del I. P. N. \\ 
Apdo. 14-740, CDMX, M\'{e}xico}

\maketitle

\begin{abstract}
We show that there is no superradiance for the Dirac field in the rotating BTZ black hole if the field vanishes at infinity. Then we outline the calculation of the expression for the renormalized energy-momentum tensor,  the effective action as well as the heat kernel for the Dirac field for the  BTZ black hole.  Finally we point out how to construct the Hartle-Hawking-Israel state for the real scalar field in the non-rotating BTZ black hole in two and three dimensions. 
\end{abstract}
\vspace{1cm}

\vspace{0.5cm}


\newpage

\section{Introduction}

The study of classical and quantum fields in black holes has been an interesting activity since the discovery of these black objects. For its relevance with reality perhaps these studies are more appealing in four dimensions, being the Kerr black hole the more relevant for these purposes. The study of quantum fields on this spacetime dates back to 1974 \cite{un74}. However from the pure theoretical point of view studying classical and quantum fields in black holes in other dimensions is a very important activity. In particular the BTZ black hole \cite{ba92}, \cite{ba93} is one stimulating object to study since shares some features with the Kerr black hole. The purpose of the present work is to go in this direction and we will address several aspects of classical and quantum fields in the BTZ black hole.

As our preferred fields will be the real scalar field and the Dirac field. In these studies we take advantage of known results regarding the solution of the Klein-Gordon operator and the Dirac operator in the rotating BTZ black hole. Armed with these results we are able to obtain interesting and closed further results. 

The organization of the paper is the following: in section 2 we show there is no superradiance for the Dirac field  in the rotating BTZ black hole if Dirichlet boundary conditions are imposed at infinity. In section 3 we obtain the renormalized energy-momentum tensor and we calculate the heat kernel and effective action for the Dirac field in the same background. In section 4 we outline the construction of the Hartle-Hawking-Israel state for the real scalar field in the non-rotating BTZ black hole in two and three dimensions.  Finally, in section 5 we give our conclusions. In a first appendix, we make some comments on the Dirac equation in three dimensional AdS and supersymmetry. Also,  for sake of clarity,  we add an appendix explaining the construction of the Hartle-Hawking-Israel state.

\section{No superradiance for the Dirac field}

In this section we will proof that there is no superradiance for the Di
rac field if we impose Dirichlet boundary conditions at infinity.

First we will write the Dirac equation in curved spacetime. This equation is
\begin{equation}
\left[\gamma^{\nu}\left(\partial_{\nu}-\Gamma_{\nu}\right)+m\right]\psi=0,
\end{equation}
where $\gamma^{\nu}$ are the Dirac matrices, $\Gamma_{\nu}$ is the spin connection and $m$ is the mass of the field. The spin connection is given by the following expression \cite{po10}
\begin{equation}
\Gamma_{\nu}=-\frac{1}{4}\gamma^{\mu}(\partial_{\nu}\gamma_{\mu}-\gamma_{\delta}\Gamma^{\delta}_{\mu\nu}),
\end{equation}
where the Dirac matrices satisfy
\begin{equation}
\gamma^{\nu}\gamma^{\mu}+\gamma^{\mu}\gamma^{\nu}=2g^{\nu\mu},
\end{equation}
with $g^{\nu\mu}$ the metric of the spacetime in consideration, and $\Gamma^{\delta}_{\mu\nu}$ the usual Christoffel symbols.

The metric we will consider is the BTZ metric \cite{ba92}, \cite{ba93}
\begin{equation}\label{scha}
ds^2=g_{tt}dt^2+g_{\phi\phi}d\phi^2+2g_{t\phi}dtd\phi+g_{rr}dr^2
\end{equation}
with
\begin{equation}
g_{tt}=M-\frac{r^2}{l^2},\hspace{0.2cm}g_{t\phi}=-\frac{J}{2},\hspace{0.2cm}g_{\phi\phi}=r^2,\hspace{0.2cm}g_{rr}=\left(-M+\frac{J^2}{4r^2}+\frac{r^2}{l^2}\right)^{-1},
\end{equation}
where $M$ is the mass of the black hole and $J$ the angular momentum. $l^2=-\frac{1}{\Lambda}$ with $\Lambda$ the cosmological constant. The outer and inner horizons are defined as
\begin{equation}
r_{\pm}^2=\frac{Ml^2}{2}\left(1\pm\sqrt{1-\frac{J^2}{M^2l^2}}\right).
\end{equation}

Now we write the metric of the BTZ black hole in a different form. We follow \cite{ar99} in writing the Dirac equation for the BTZ black hole. The metric is given by
\begin{equation}\label{btz}
ds^2=-\frac{\Delta^2}{l^2\rho^2}dt^2+\frac{l^2\rho^2}{\Delta^2}d\rho^2+\rho^2\left(d\phi-\frac{\rho_{+}\rho_{-}}{l\rho^2}dt\right)^2,
\end{equation}
where $\rho=r$ is the radial coordinate and 
\begin{equation}
\Delta^2=(\rho^2-\rho_{+}^2)(\rho^2-\rho_{-}^2).
\end{equation}
Doing $\rho^2=\rho_{+}^2\cosh^2\mu-\rho_{-}^2\sinh^2\mu$ then\footnote{In these coordinates the horizon is at $\mu=0$ and infinity at $\mu=\infty.$}
\begin{equation}
ds^2=-\sinh^2\mu\left(\rho_{+}\frac{dt}{l}-\rho_{-}d\phi\right)^2+l^2 d \mu^2+\cosh^2\mu\left(-\rho_{-}\frac{dt}{l}+\rho_{+}d\phi\right)^2.
\end{equation}
Now we introduce the coordinates 
\begin{equation}
x^{+}=\rho_{+}\frac{t}{l}-\rho_{-}\phi\hspace{0.5cm}x^{-}=-\rho_{-}\frac{t}{l}+\rho_{+}\phi,
\end{equation}
then
\begin{equation}
ds^2=-\sinh^2\mu dx^{+2}+ l^2 d \mu^2+\cosh^2\mu dx^{-2}.
\end{equation}
Choosing 
\begin{equation}
e^{0}_{+}=\sinh\mu\hspace{0.5cm}e^{2}_{-}=\cosh\mu\hspace{0.5cm}e^1_{\mu}=l,
\end{equation}
we have
\begin{equation}
\Gamma_{+}=\frac{1}{2l}\widetilde{\gamma}_{2}\widetilde{\gamma}_{0}\cosh\mu\hspace{0.5cm}\Gamma_{-}=-\frac{1}{2l}\widetilde{\gamma}_{1}\widetilde{\gamma}_{2}\sinh\mu\hspace{0.5cm}\Gamma_{\mu}=0,
\end{equation}
where
\begin{equation}
\widetilde{\gamma}_{0}=i\sigma_{2}\hspace{0.5cm}\widetilde{\gamma}_{1}=\sigma_{1}\hspace{0.5cm}\widetilde{\gamma}_{2}=\sigma_{3}
\end{equation}
with $\sigma_{1}, \sigma_{2}, \sigma_{3}$ the usual Pauli matrices. Using the obtained spin connection coefficients we have the Dirac equation in (\ref{btz}) 
\begin{equation}
\left(-\frac{1}{\sinh\mu}\widetilde{\gamma}_{0}\partial_{+}+\frac{1}{\cosh\mu}\widetilde{\gamma}_{2}\partial_{-}+\frac{\widetilde{\gamma}_{1}}{l}\partial_{\mu}+\frac{1}{2l}\left(\frac{\cosh\mu}{\sinh\mu}+\frac{\sinh\mu}{\cosh\mu}\right)\widetilde{\gamma}_{1}+m\right)\psi=0.
\end{equation}
Writing the spinor as
\begin{equation}
\psi=\frac{1}{\sqrt{\sinh\mu\cosh\mu}}e^{-i(k^+x^{+}+k^-x^{-})}
\left(\begin{array}{c}
	\psi_{1}\\
	\psi_{2}
\end{array}\right),
\end{equation}
where $k^+=(\omega-\widetilde{m}\Omega)/(2\pi lT_{H})$, $k^-=(\rho_{-}\omega-\rho_{+}\widetilde{m}/l)/(2\pi l\rho_{+}T_{H})$, $T_{H}=(\rho_{+}^2-\rho_{-}^2)/2\pi l^2\rho_{+}$ and $\Omega=J/2\rho_{+}^2$. We find that
\begin{eqnarray}
\left(\frac{ik^+}{\sinh\mu}-\frac{ik^-}{\cosh\mu}\right)\frac{\psi_{2}}{\sqrt{\sinh\mu\cosh\mu}} & = & -\left(\frac{1}{2l}\left(\frac{\cosh\mu}{\sinh\mu}+\frac{\sinh\mu}{\cosh\mu}\right)+\right.\nonumber\\
&  & \left.+ m+\frac{1}{l}\partial_{\mu}\right)\frac{\psi_{1}}{\sqrt{\sinh\mu\cosh\mu}},
\end{eqnarray}
\begin{eqnarray}
-\left(\frac{ik^+}{\sinh\mu}+\frac{ik^-}{\cosh\mu}\right)\frac{\psi_{1}}{\sqrt{\sinh\mu\cosh\mu}} & = & -\left(-\frac{1}{2l}\left(\frac{\cosh\mu}{\sinh\mu}+\frac{\sinh\mu}{\cosh\mu}\right)+\right.\nonumber\\
&  & \left.+ m-\frac{1}{l}\partial_{\mu}\right)\frac{\psi_{2}}{\sqrt{\sinh\mu\cosh\mu}}.
\end{eqnarray}
Further simplification gives 
\begin{equation}
\left(d_{\mu}-\frac{ilk^+}{\sinh\mu}\right)\psi_{2}=-\left(ml-\frac{ilk^-}{\cosh\mu}\right)\psi_{1},
\end{equation}
\begin{equation}
\left(d_{\mu}+\frac{ilk^+}{\sinh\mu}\right)\psi_{1}=-\left(ml+\frac{ilk^-}{\cosh\mu}\right)\psi_{2}.
\end{equation}
Now we introduce another base of functions
\begin{equation}
\psi_{1}+\psi_{2}=(1-\tanh^2\mu)^{-1/4}\sqrt{1+\tanh\mu}(\psi_{1}'+\psi_{2}')
\end{equation}
\begin{equation}
\psi_{1}-\psi_{2}=(1-\tanh^2\mu)^{-1/4}\sqrt{1-\tanh\mu}(\psi_{1}'-\psi_{2}').
\end{equation}
Making $y=\tanh\mu$ then we have\footnote{In this coordinate the horizon is at $y=0$ and infinity at $y=1$.}
\begin{equation}\label{eq21}
(1-y^2)d_{y}\psi_{2}'-il\left(\frac{k^{+}}{y}+k^{-}y\right)\psi_{2}'=-\left(ml+\frac{1}{2}-il(k^{+}+k^{-})\right)\psi_{1}'
\end{equation}
\begin{equation}\label{eq22}
(1-y^2)d_{y}\psi_{1}'+il\left(\frac{k^{+}}{y}+k^{-}y\right)\psi_{1}'=-\left(ml+\frac{1}{2}+il(k^{+}+k^{-})\right)\psi_{2}'.
\end{equation}
These equations are separable. Solving for example for $\psi_{1}'$ we have\footnote{In this equation $d$ means $d_{z}$.}
\begin{eqnarray}
& &z(1-z)d^2\psi_{1}'+\frac{1}{2}(1-3z)d\psi_{1}'+\nonumber\\
& &+\frac{1}{4}\left(\frac{-ilk^{+}+l^2{k^{+}}^2}{z}+ilk^{-}-l^2{k^{-}}^2-\frac{(ml+1/2)^2}{1-z}\right)\psi_{1}'=0,
\end{eqnarray}
where\footnote{The horizon is at $z=0$ and infinity is at $z=1$.} $y^2=z$. This is a hypergeometric equation whose solution  is
\begin{eqnarray}
\psi_{1}'& = &\frac{k_{1}}{\sqrt{1-z}}z^{-i (lk^{+})/2}(z-1)^{(1-2lm)/4}{_{2}F_{1}(a,b;c;z)}\nonumber\\
         &   & +\frac{k_{2}ie^{\pi(-lk^{+})}}{\sqrt{1-z}}z^{(1+ilk^{+})/2}(z-1)^{(1-2lm)/4}{_{2}F_{1}(d,e;f;z)},
\end{eqnarray}
where $a=-(2lk^{+}-2lk^{-}-2iml+i)/4i$, $b=-(2lk^{+}+2lk^{-}-i(2ml+1))/4i$, $c=1/2-ilk^{+}$, $d=(2ilk^{+}-2ilk^{-}-2ml+1)/4$, $e=(2ilk^{+}+2ilk^{-}-2lm+3)/4$, $f=ilk^{+}+3/2$ and $k_{1}$ and $k_{2}$ are constants. Here we see that if $-1/2\geq ml$ then we can make $\psi_{1}'=0$ at infinity. This demands that $l$ to be negative. The bound that we have just found seems similar to the one found in AdS for the Dirac field \cite{ba08}. We also comment that our solutions include the solution given by Dasgupta \cite{ar99} and it is obtained when $ml=1/2$. Also it is clear that we can not have a well behaved massless spinor.

Let us take $k_{1}=0$. Then the solution is
\begin{equation}
\psi_{1}'=\frac{1}{\sqrt{1-z}}z^{(1+ilk^{+})/2}(z-1)^{(1-2lm)/4}{_{2}F_{1}(d,e;f;z)}.
\end{equation}
To obtain $\psi_{2}'$ we have to differentiate the previous function, see (\ref{eq22})
\begin{equation}
\frac{d\psi_{1}'}{dz}=\alpha z^{\alpha-1}(1-z)^{\beta}F-\beta z^{\alpha}(1-z)^{\beta-1}F+z^{\alpha}(1-z)^{\beta}\frac{ab}{c}G,
\end{equation}
where $\alpha=(1+ilk^{+})/2$, $\beta=-(1+2lm)/4$ and $G=F(d+1,e+1;f+1;z)$. If we want $\psi_{2}'=0$ at infinity then we have one more restriction in the parameter $\beta$: $\beta-1>0$.

From the above considerations we see that $\psi_{1}'$ and $\psi_{2}'$ are zero at infinity. Now let us see what happens with the original spinor. We have that
\begin{equation}
(1-\tanh^2\mu)^{-1/4}\sqrt{1+\tanh\mu}=e^{\mu/2}.
\end{equation}
Then
\begin{equation}
\psi=\sqrt{\frac{2}{\sinh 2\mu}}e^{-i(k^{+}x^{+}+k^{-}x^{-})}
\left(\begin{array}{c}
\cosh\mu/2\psi_{1}'+\sinh\mu/2\psi_{2}'\\
\sinh\mu/2\psi_{1}'+\cosh\mu/2\psi_{2}'
\end{array}\right).
\end{equation}
Then clearly Dirichlet boundary conditions at infinity for the primes imply Dirichlet boundary conditions for the original spinor. Now let us see what happens at the horizon. When we approach the horizon $\mu\rightarrow 0$ then we have infinities, and the black hole seems to be unstable, unless $\psi_{1}'=\psi_{2}'=0$ at the horizon. So it seems that the physics of the problem force us to put some sort of brick wall and then there is no superradiance. 
We point out that these boundary conditions are quite natural, since the BTZ black hole is asymptotically AdS spacetime, and it has been shown \cite{Avis1978} that a well-defined quantization scheme can be set up in  AdS spacetime with these boundary conditions.

One can ask if the non-existence of superradiance shown above  depends on the ansatz for the spinor. In order to answer this question let us take some limits in the Dirac equation. If we make $\mu\rightarrow\infty$ then we obtain
\begin{equation}
\partial_{\mu}g_{2,1}+g_{2,1}+mlg_{1,2}=0,
\end{equation} 
where we have written the spinor as $\psi=e^{-i(k^{+}x^{+}+k^{-}x^{-})}(g_{1},g_{2})^{T}$. Solving for $g_{1}$ we find that
\begin{equation}
g_{1}(\mu)=C_{1}e^{-(ml+1)\mu}+C_{2}e^{(ml-1)\mu},
\end{equation}
hence there are bounds for $m$ depending on the boundary conditions of the field at infinity. Now let us see what happens when $\mu\rightarrow 0$. In this case we find that
\begin{equation}
(ik^{+}\widetilde{\gamma}_{0}+\frac{1}{2l}\widetilde{\gamma}_{1})\psi=0.
\end{equation}
This implies $g_{1,2}=0$ and for example the Dirac current is zero at the horizon and then there is no superradiance. A similar result for the scalar field was obtained in \cite{or12} and a related result was obtained in \cite{da17} with Robin boundary conditions at infinity. From the above analysis we see that the non-existence of superradiance is quite generic, however the analysis of this paragraph should be taken with care since we are over simplifying the problem, in the next paragraph we will prove that there is no superradiance. 

In order to prove the non-existence of superradiance we have to express our solution which vanishes at infinity, where $z=1$, around $1-z$ using the linear transformations of the hypergeometric functions. Explicitly we will use the following relation \cite{abra}
\begin{eqnarray}
F(a,b;c;z) & = & \frac{\Gamma(c)\Gamma(c-a-b)}{\Gamma(c-a)\Gamma(c-b)}F(a,b;a+b-c+1;1-z)\\
           & + & (1-z)^{c-a-b}\frac{\Gamma(c)\Gamma(a+b-c)}{\Gamma(a)\Gamma(b)}F(c-a,c-b;c-a-b+1;1-z).\nonumber
\end{eqnarray}
If now we let $z\rightarrow 0$ then we obtain that the first component of the spinor goes like $z^{\frac{1}{4}}$ times a finite number, hence at the horizon $z=\mu=0$ we have that this component vanishes, however the second component does not vanish at $\mu=0$  and if we calculate the radial component of the Dirac current \cite{ca13} we obtain\footnote{Here the components of the spinor are the ones without the factor $\frac{1}{\sqrt{\sinh\mu\cosh\mu}}$. However for saying if there is superradiance or not the spinor without this term in enough.} 
\begin{equation}\label{eq32}
j^{\mu}=l(\left|\psi_{1}^{*}\right|^2-\left|\psi_{2}^{*}\right|^2)
\end{equation}
which implies a flux towards the black hole and hence we have no superradiance. Hence for our choice of boundary conditions at infinity there is no superradiance however (\ref{eq32}) indicates that it might be possible that for other boundary conditions there could be superradiance, we leave this question for a future work.
See also Appendix A.

In the following sections we address miscellaneous  aspects related to quantum fields in BTZ spacetime; these remarks are complementary and have also a didactic purpose, pointing out some details that in most papers are omitted. 
These are related to the energy-momentum tensor, effective action and heat kernel of the Dirac equation, while  section 4 regards the calculation of the  Hartle-Hawking state. 

\section{Aspects of the Dirac Equation in BTZ black hole spacetime}

In this section we obtain the renormalized energy-momentum tensor and we calculate the heat kernel and effective action for the Dirac field in the BTZ background.

\subsection{Towards the energy-momentum tensor for Dirac}

The classical energy-momentum tensor for the Dirac field is \cite{ca13}
\begin{equation}
T_{\mu\nu}=\frac{i}{2}\left[\overline{\psi}\gamma_{(\mu}\nabla_{\nu)}\psi-(\nabla_{(\mu}\overline{\psi})\gamma_{\nu)}\psi\right].
\end{equation}
Hence the expectation value of the quantum version of this tensor in the appropriate state is given by \cite{am13}
\begin{equation}
\left\langle T_{\mu\nu}\right\rangle=\lim_{x'\rightarrow x}\textnormal{tr}\left\{\frac{i}{2}\left[\gamma_{(\mu}\nabla_{\nu)}-\gamma_{(\mu'}\nabla_{\nu')}\right]G^{D}_{R}(x,x')\right\},
\end{equation}
where $\nabla_{\mu}$ is the spinor covariant derivative. In our problem $G^{D}_{R}$ is the renormalized Feynman propagator associated with the Hartle-Hawking-Israel state for the rotating BTZ black hole. For the massless case, without renormalization, this propagator is known and it is given by \cite{hy95}
\begin{equation}
G^{D}(x,x')=\sum_{n=-\infty}^{\infty}G_{A}(x,x'_{n}),
\end{equation}
where $G_{A}$ is the propagator in AdS and $x'=(t',r',\phi')$ and $x'_{n}=(t',r',\phi'+2\pi n)$. The renormalization of this propagator is not known and we will take an intuitive approach. Assuming that the untraviolet divergences are the same for the scalar and the Dirac field then we can use the results given in \cite{de08} in our problem, and then the renormalized Feynman propagator is given by
\begin{equation}
G^{D}_{R}(x,x')=G^{D}(x,x')-G^{F}_{sing}(x,x'),
\end{equation}
where
\begin{equation}
G^{F}_{sing}(x,x')=\frac{i}{4\sqrt{2}\pi}\left(\frac{U(x,x')}{[\sigma(x,x')+i\epsilon]^{1/2}}\right).
\end{equation}
For the scalar case $U(x,x')$ is a scalar, we will assume that for the Dirac field $U(x,x')$ is a spinor. This spinor can be calculated recursively up to the required order \cite{de08}. $\sigma(x,x')$ is the half geodesic distance between $x$ and $x'$.

The first terms of the expansion of the singular propagator can be written in this case. We have
\begin{equation}
U=U_{0}+U_{1}\sigma+O(\sigma^2)
\end{equation}
with
\begin{equation}
U_{0}=1-\frac{2}{l^2}\sigma,
\end{equation} 
\begin{equation}
U_{1}=m^2-\frac{6}{l^2}(\xi-1/6).
\end{equation}
For a massless conformal field the last term vanishes. In order to calculate the energy-momentum tensor we have to know the expression for $G^{D}(x,x')$, however as the reader can see this expression can not be expressed in a closed form. Hence we leave at this point the calculation of the energy-momentum tensor. Just for completeness we give the expression of $\sigma$ in terms of BTZ coordinates
\begin{eqnarray}
\sigma(x,x') & = & \frac{1}{2}\eta_{\mu\nu}(\xi^\mu-\xi'^\mu)(\xi^\nu-\xi'^\nu)\nonumber\\
             & = & \frac{1}{2}\left(B+B'-A-A'-\sqrt{BB'}\cosh(t-t')+\sqrt{AA'}\cosh(\phi-\phi')\right),\nonumber
\end{eqnarray}
where \cite{or11}
\begin{equation}
A=\frac{l^2}{r^2_{+}}r^2 
\end{equation}
and
\begin{equation}
B=l^2\left(\frac{r^2-r^2_{+}}{r^2_{+}}\right).
\end{equation}

Further steps might  be done by numerical calculation that is beyond our present purposes.

\subsection{Some words about vacuum polarization}

For the scalar field in the BTZ black hole there is polarization of the vacuum \cite{sh941}, \cite{sh942}. This polarization goes as $1/r$ and then vanishes at infinity. However for the same field but in pure AdS there is also polarization which is, as expected, constant \cite{ke15} and then does not vanish at infinity. Since the BTZ black hole is locally AdS we can conclude that the polarization of the vacuum is a global effect. This is not something completely new since the definition of vacuum is a global definition. If we could calculate the energy-momentum tensor for Dirac it would be interesting if we obtain the same vacuum polarization as with the scalar field.  

\subsection{Effective action and heat kernel} 

We are interested in calculating the effective action and the heat kernel for the Dirac field in the BTZ black hole. In this subsection we will do this using an asymptotic expansion.

The effective action for the Dirac field is given as \cite{to09}
\begin{equation}
W=-\frac{i\hbar}{2}\ln \left[\textnormal{det}(\textsl{l}D)\right]^2
\end{equation}
where $\textsl{l}$ is length and $D$ is the Dirac operator. The last expression can be written as
\begin{equation}
W=-\frac{i\hbar}{2}\ln\textnormal{det}\left[\textsl{l}^2\left(\nabla_{\mu}\nabla^{\mu}+\frac{1}{4}R+m^2\right)\right],
\end{equation}
where $\nabla_{\mu}$ is the spinorial covariant derivative and we have used the identity \cite{to09}, \cite{ca92}
\begin{equation}
\left(\gamma^{\mu}\nabla_{\mu}\right)^2+m^2=\nabla_{\mu}\nabla^{\mu}+\frac{1}{4}R+m^2.
\end{equation}
The utility of having expressed the effective action as function of a second order differential operator is that for this kind of operators it is well known how to write the divergences of the effective action. For example for an operator
\begin{equation}
D_{x}=g^{\mu\nu}(x)\nabla_{\mu}\nabla_{\nu}+Q(x)
\end{equation}
the heat kernel has the following expansion \cite{to09}
\begin{equation}
K(\tau;x,x)\approx i(4\pi i\tau)^{-n/2}\sum_{k=0}^{\infty}{(i\tau)^kE_{k}(x)},
\end{equation}
where 
\begin{equation}
E_{0}(x)=I
\end{equation}
\begin{equation}
E_{1}(x)=\frac{1}{6}RI-Q
\end{equation}
\begin{eqnarray}
E_{2}(x) & = & \left(-\frac{1}{30}\nabla_{\mu}\nabla^{\mu}R+\frac{1}{72}R^2-\frac{1}{180}R^{\mu\nu}R_{\mu\nu}+\frac{1}{180}R^{\mu\nu\delta\sigma}R_{\mu\nu\delta\sigma}\right)I\nonumber\\
& + & \frac{1}{12}W^{\mu\nu}W_{\mu\nu}+\frac{1}{2}Q^2-\frac{1}{6}RQ+\frac{1}{6}\nabla_{\mu}\nabla^{\mu}Q
\end{eqnarray}
with
\begin{equation}
W_{\mu\nu}=[\nabla_{\mu},\nabla_{\nu}].
\end{equation}

The effective action is given in terms of the heat kernel as
\begin{equation}
W=-\frac{i\hbar}{2}\int{dv_{x}}\int_{0}^{\infty}{\frac{d\tau}{\tau}\textnormal{tr}K(\tau;x,x)}.
\end{equation}
It is usually the case that this expression has divergences, due to the lower limit of the integral on $\tau$. Hence these divergences are isolated as follows. If we denote \textbf{divp} as the divergence part of any expression then
\begin{equation}
\textbf{divp}W=-\frac{i\hbar}{2}\int{dv_{x}}\textbf{divp}\int_{0}^{\tau_{0}}{\frac{d\tau}{\tau}\textnormal{tr}K(\tau;x,x)},
\end{equation} 
where $\tau_{0}$ is a small constant. Now using the expansion of the heat kernel for small proper time we have
\begin{equation}
\textbf{divp}W=\frac{i\hbar}{2}(4\pi)^{-n/2}\int{dv_{x}}\sum_{k=0}^{\infty}{\textnormal{tr}E_{k}(x)\textbf{divp}\int_{0}^{\tau_{0}}{d\tau(i\tau)^{k-1-n/2}}}.
\end{equation}
In this expression the divergences are clearly isolated. Now we regularize this expression by a cut-off and dimensional regularization. Since we are interested in the BTZ black hole we will give the expressions for three dimensions and refer the reader to \cite{to09}. For three dimensions we have that
\begin{equation}
\textbf{divp}W=-\frac{\hbar}{2}(4\pi)^{-3/2}\int{dv_{x}}\sum_{k=0}^{1}\textnormal{tr}E_{k}(x)(i\textsl{l}^2)^{k-3/2}\frac{\tau_{c}^{k-3/2}}{k-\frac{3}{2}},
\end{equation}
where $\tau_{c}$ is a small proper time. In the dimensional regularization there is no divergent part in the effective action.
\section{The Hartle-Hawking-Israel state in the BTZ black hole}

It is well known \cite{fro97} that in the Schwarzschild black hole there exists for the real scalar field three possible vacuum known as the Hartle-Hawking-Israel \cite{haho76}, \cite{is76}, the Boulware \cite{bo75} and the Unruh \cite{un76} vacuum. Some time ago \cite{or94} it was found that the global quantization in AdS induces the Hartle-Hawking-Israel state in the BTZ black hole. In this section we will construct this state using the solutions of the Klein-Gordon equation in the BTZ black hole. 

In this section we show how to contruct the Hartle-Hawking-Israel state for the two and three dimensional BTZ black hole. For the three dimensional case this state has been given by \cite{or94} but by means of the immersion of the BTZ black hole in AdS. In this section we construct this state by using the solutions of the Klein-Gordon equation in the BTZ black hole without viewing it as immersed in AdS.

\subsection{The Hartle-Hawking-Israel state in two dimensions}

The two dimensional BTZ black hole can be considered as a solution of the Jackiw-Teitelboim theory \cite{mCasMi95}, its metric is given by\footnote{This metric can also be considered as a dimensional reduction of the BTZ metric in three dimensions \cite{aAchumOrt93}.}
\begin{equation}
ds^2=-\left(-M+\frac{r^2}{l^2}\right)dt^2+\frac{dr^2}{\left(-M+\frac{r^2}{l^2}\right)},
\end{equation}
where $M$ and $l$ are parameters of the theory. This metric has a Kruskal like extension \cite{or11} and we can introduce a tortoise like coordinate \cite{or11} $r^{*}$ in such a way that the metric takes the form
\begin{equation}\label{eq.1}
ds^2=\left(-M+\frac{r^2}{l^2}\right)(dt^2+dx^2),
\end{equation}
where we have made $r^{*}=x$ and $r$ is an implicit function of $x$. The coordinate $x$ is $-\infty$ at the horizon and $0$ at infinity.

Now we consider the massless conformally coupled real scalar field with equation of motion
\begin{equation}\label{eq.2}
\nabla_{\mu}\nabla^{\mu}\phi=0
\end{equation}
in the metric (\ref{eq.1}). The equation (\ref{eq.2}) is such that it reduces to
\begin{equation}
(-\partial^2_{t}+\partial^2_{x})\phi=0.
\end{equation}
If we assume a harmonic dependence in time and impose Dirichlet boundary conditions at infinity\footnote{We have to impose boundary conditions at infinity since the BTZ black hole has a time-like boundary at infinity.} then we have the solutions
\begin{equation}\label{eq.3}
\phi(t,x)=\frac{1}{\sqrt{\pi\omega}}e^{-i\omega t}\sin(\omega x).
\end{equation}
These solutions are normalized with respect to the Klein-Gordon inner product \footnote{If we choose Neumann boundary conditions then the normalized solutions are with cosine instead of sine.}.

The Kruskal coordinates $T$ and $R$ are given in
    terms of $t$ and $r^{\ast}$ as \cite{or11}:
    
    In $\mathcal{R}_{K}$
    \begin{equation}\label{E:b28}
    T=e^{\frac{r_{+}}{l^{2}}r^{\ast}}\sinh\left(\frac{r_{+}}{l^{2}}t\right)\qquad
    R=e^{\frac{r_{+}}{l^{2}}r^{\ast}}\cosh\left(\frac{r_{+}}{l^{2}}t\right).
    \end{equation}
    In $\mathcal{L}_{K}$
    \begin{equation}\label{E:b28a}
    T=-e^{\frac{r_{+}}{l^{2}}r^{\ast}}\sinh\left(\frac{r_{+}}{l^{2}}t\right)\qquad
    R=-e^{\frac{r_{+}}{l^{2}}r^{\ast}}\cosh\left(\frac{r_{+}}{l^{2}}t\right).
    \end{equation}
    In $\mathcal{F}_{K}$
    \begin{equation}\label{E:b28b}
    T=e^{\frac{r_{+}}{l^{2}}r^{\ast}}\cosh\left(\frac{r_{+}}{l^{2}}t\right)\qquad
    R=e^{\frac{r_{+}}{l^{2}}r^{\ast}}\sinh\left(\frac{r_{+}}{l^{2}}t\right).
    \end{equation}
    In $\mathcal{P}_{K}$
    \begin{equation}\label{E:b28c}
    T=-e^{\frac{r_{+}}{l^{2}}r^{\ast}}\cosh\left(\frac{r_{+}}{l^{2}}t\right)\qquad
    R=-e^{\frac{r_{+}}{l^{2}}r^{\ast}}\sinh\left(\frac{r_{+}}{l^{2}}t\right).
    \end{equation}
    Here $\mathcal{R}_{K}$, $\mathcal{L}_{K}$, $\mathcal{F}_{K}$ and $\mathcal{P}_{K}$ are the right, left, future and past regions of the Kruskal spacetime. From these expressions we see that the $(T,R)$ and $(t,r^{\ast})$ coordinates
    are related as the Minkowski and Rindler coordinates in flat
    spacetime are.
    
    Following, for example, \cite{cKr10} or the description of the construction of Israel given in the appendix A we can introduce Rindler like modes in $\mathcal{R}_{K}$ and $\mathcal{L}_{K}$. These modes will be given in terms of (\ref{eq.3}). It is clear from the expressions of the normalized modes and the form of the Kruskal coordinates in terms of $t$ and $r^{\ast}$ that every step given in \cite{cKr10} applies to the present case and then we can obtain the Hartle-Hawking-Israel state $\left|K\right\rangle$ which would be the analogous of the Minkowski vacuum in terms of Rindler vacuum $\left|R\right\rangle$
\begin{equation}
\left|K\right\rangle=\texttt{exp}\left\{\sum_{\omega}{\left(-\ln\cosh\phi_{\omega}+(\tanh\phi_{\omega})a_{\omega}^{(+)\dagger}a_{\omega}^{(-)\dagger}\right)}\right\}\left|R\right\rangle,
\end{equation}		
where $a_{\omega}^{(+)\dagger}$ and $a_{\omega}^{(-)\dagger}$ are the creation operators, see also appendix A, associated with the modes (\ref{eq.3}) and with its partners on the other wedge respectively. The state we obtain is analogous to the one obtained by \cite{is76} however in our case we have given an explicit expression for the functions which define the Rindler like quantization whereas Israel just assumed the existence of such functions. Also we point out that our arguments clarify the way the Hartle-Hawking-Israel state is obtained from first principles.

\subsection{The Hartle-Hawking-Israel state in three dimensions}

We assume the field satisfies the following equation
\begin{equation}
(\nabla_{\mu}\nabla^{\mu}-\xi R-m^2)\varphi=0,
\end{equation}
where $\xi$ is a coupling constant, $R$ is the Ricci scalar and $m$ can be consider the mass of the field when it is quantized.

Now we assume the field has the following form $\varphi=e^{-i\omega t+i m\phi}R(r)$. Once we do this the Klein-Gordon equation in the metric (\ref{scha}) reduces to
\begin{equation}
\left(g_{rr}(\omega-\frac{J}{2r^2}m)^2-\frac{m^2}{r^2}+\frac{1}{r}\partial_{r}\frac{r}{g_{rr}}\partial_{r}-\widetilde{m}^2\right)R(r)=0,
\end{equation}
where $\widetilde{m}^2=m^2-\frac{6\xi}{l^2}$. Now let us introduce a new variable \cite{ku08} $z$ and a function $F(z)$ as
\begin{equation}
z=\frac{r^2-r_{+}^2}{r^2-r_{-}^2},\hspace{0.3cm}F(z)=z^{i\alpha}(1-z)^{-\beta}R(z),
\end{equation}
then we obtain the hypergeometric differential equation 
\begin{equation}
z(1-z)\frac{d^2F}{dz^2}+(c-(1+a+b)z)\frac{dF}{dz}-abF=0,
\end{equation}
with the following parameters
\begin{equation}
a=\beta-i\frac{l^2}{2(r_{+}+r_{-})}\left(\omega+\frac{m}{l}\right)\hspace{0.2cm}b=\beta-i\frac{l^2}{2(r_{+}-r_{-})}\left(\omega-\frac{m}{l}\right)\hspace{0.2cm}c=1-2i\alpha,
\end{equation}
with
\begin{equation}
\alpha=\frac{l^2r_{+}}{2(r_{+}^2-r_{-}^2)}(\omega-\Omega_{H}m)\hspace{0.3cm}\beta=\frac{1-\sqrt{1+l^2\widetilde{m}^2}}{2},
\end{equation}
where $\Omega_{H}$ is the angular velocity of the horizon. 

Following \cite{ku08} we impose Dirichlet boundary conditions at infinity, which give us the following solution
\begin{equation}
R_{\infty}=\frac{z^{-i\alpha}(1-z)^{\beta}(1-z)^{c-a-b}}{\Gamma(c-a-b+1)}F(c-a,c-b,c-a-b+1;1-z).
\end{equation}
Near the horizon this solution can be expressed as
\begin{equation}
R_{\infty}=\frac{\Gamma(1-c)}{\Gamma(1-a)\Gamma(1-b)}R_{r_{+}in}+\frac{\Gamma(c-1)}{\Gamma(c-a)\Gamma(c-b)}R_{r_{+}out},
\end{equation}
where in terms of the hypergeometric function we have
\begin{equation}
R_{r_{+}in}=z^{-i\alpha}(1-z)^{\beta}F(a,b,c;z),
\end{equation}
\begin{equation}
R_{r_{+}out}=z^{i\alpha}(1-z)^{\beta}F(1+b-c,1+a-c,2-c;z).
\end{equation}
The approximate expression of these functions near the horizon is
\begin{equation}
R_{r_{+}in}\approx exp(-i(\omega-\Omega_{H}m)r^{*}-i\alpha_{0}(\omega)),
\end{equation}
\begin{equation}
R_{r_{+}out}\approx exp(i(\omega-\Omega_{H}m)r^{*}+i\alpha_{0}(\omega)),
\end{equation}
where $r^{*}$ is the tortoise coordinate and 
\begin{equation}
\alpha_{0}(\omega)=\frac{l^2r_{+}(\omega-\Omega_{H}m)}{2(r_{+}^2-r_{-}^2)}\texttt{ln}\frac{4r_{+}^2}{r_{+}^2-r_{-}^2}.
\end{equation}
At this point Kuwata \textit{et.} \textit{al.} \cite{ku08} choose boundary conditions at the horizon. We will follow a different path and point out that the Hartle-Hawking-Israel state can be constructed following \cite{is76} since if we write the modes near the horizon in term of the Kruskal coordinates\footnote{These coordinates are null Kruskal coordinates which in terms of the usual Kruskal coordinates, $T$ and $R$, are defined as $U=T-R$ and $V=T+R$. Here $T$ and $R$ are defined in terms of the Schwarzschild time $t$ and the tortoise coordinate $r^*$ as defined in the precedent section in two dimensions.} $V$ and $U$ \cite{or11} we have
\begin{equation}
\varphi_{in}\approx exp(-i\frac{\omega}{\kappa}\textnormal{ln}V),
\end{equation}
\begin{equation}
\varphi_{out}\approx exp(-i\frac{\omega}{\kappa}\textnormal{ln}(-U)),
\end{equation}
where $\kappa$ is the surface gravity. Using this expressions for the modes it is clear that we can construct the Hartle-Hawking-Israel state following \cite{is76}. For more details see the paper by Israel (see also Appendix B). Hence we conclude that the Hartle-Hawking-Israel state can be constructed for the non-rotating BTZ black hole. It is interesting that if we allow rotation then the modes do not have the simple expressions just mentioned, this can be an indication that it is not possible to construct the Hartle-Hawking-Israel state for the rotating BTZ black hole. If this is true we would have indications that a theorem like the Kay-Wald theorem for Kerr\footnote{What this theorem says is tantamount, in the present context, to say that there not exist the Hartle-Hawking-Israel state for the real scalar field in the Kerr metric.} \cite{ka91} holds for the rotating BTZ black hole.

\section{Conclusions}

We have discussed some interesting aspects of classical and quantum fields in the BTZ black hole. For example we have shown that there is no superradiance for the Dirac field in the BTZ black hole for Dirichlet boundary conditions at infinity. This is analogous with the Dirac field in the Kerr metric which does not have superradiance either \cite{un74}. Hence it seems that the existence of superradiance has to do with the boundary conditions at infinity as well as with the physics near the black hole.  We outlined the construction of the energy-momentum tensor for the Dirac field;  it might be possible that the corresponding calculation can be done by numerical methods. Also we have been able to write approximate expressions for the effective action and heat kernel which simplify due to the dimensionality of the BTZ black hole. Other relevant result is the construction of the Hartle-Hawking-Israel state in two and three dimensions using only the geometry of BTZ itself and without using the embedding of it in AdS.\\

\vspace{0.5cm}
\textbf{Acknowledgments}: The work of L. O. has been sponsored by CONACYT-Mexico through a postdoctoral fellowship. N. B. acknowledges partial financial support from CONACYT-Mexico through the proyect No. 284489. 

\appendix

\section{The Dirac equation in AdS and supersymmetry}

In this section we show there exist some coordinates where the Dirac equation in three dimensional AdS can be written easily and make some comments on constructing some supersymmetrical models in this spacetime.

It is well known \cite{ba99} that three dimensional AdS can be defined as the hyperboloid $-U^2-V^2+X^2+Y^2=-1/l^2$ embedded in a space with metric $ds^2=-dU^2-dV^2+dX^2+dY^2$.

Now we can introduce the so called global coordinates defined by
\begin{equation}
U=1/l\cosh\mu\sin t,\hspace{0.5cm}V=1/l\cosh\mu\cos t,
\end{equation} 
\begin{equation}
X=1/l\sinh\mu\cos\theta,\hspace{0.5cm}V=1/l\sinh\mu\sin\theta.
\end{equation} 
In these coordinates the metric reads as
\begin{equation}
ds^2=1/l^2\left(-\cosh^2\mu dt^2+d\mu^2+\sinh^2\mu d\theta^2\right).
\end{equation}
Here $0\leq\mu\leq\infty$, $0\leq\theta\leq 2\pi$ and $0\leq t\leq 2\pi$. It is clear that the expression for this metric allow us to introduce the thriads easily as we did in the BTZ black hole in the main text of this work. So we expect that an exact solution of the Dirac equation in three dimensional AdS spacetime exist.  If our expectatives are correct then we could try to construct a supersymmetrical model in three dimensional AdS spacetime as was done in \cite{sa85}. We leave this also for future work. 
   
\section{The Unruh-Israel construction}

We call the Unruh-Israel construction to the construction of the Minkowski/Krus-\
kal vacuum by using two copies of Rindler/Schwarzschild quantization. Let $f_{\omega}$ be the modes of the Klein-Gordon equation in the left or right wedge of Minkowski/Kruskal spacetime\footnote{For the case of Kruskal spacetime change right for right exterior and left for left exterior.}. Now we define $F^{(\epsilon)}_{\omega}$ to be equal to $f_{\omega}$ in the right wedge for $\epsilon=+$ and zero in the other wedge; similarly we define $F^{(\epsilon)}_{\omega}$ for $\epsilon=-$. It turns out that the combination \cite{un76}
\begin{equation}
H^{(\epsilon)}_{\omega}(x)=F^{(\epsilon)}_{\omega}(x)\cosh\phi_{\omega}+F^{(-\epsilon)}_{\omega}(x)\sinh\phi_{\omega}
\end{equation}
is analytic on the horizon, where $\tanh\phi_{\omega}=\exp(-\pi\omega/\kappa)$ and $\epsilon$ is $\pm$ depending on which wedge is $x$. The basic idea of the construction is that $F^{(\epsilon)}_{\omega}(x)$ and $H^{(\epsilon)}_{\omega}(x)$ are both complete and satisfy
\begin{equation}
(H^{(\epsilon)}_{\omega},H^{(\epsilon')}_{\omega'})=(F^{(\epsilon)}_{\omega},F^{(\epsilon')}_{\omega'})=\epsilon\delta_{\epsilon\epsilon'}\delta(\omega-\omega'),
\end{equation}
where $(,)$ is the Klein-Gordon inner product. Then both set of modes define equivalent quantizations. The quantization defined by $H^{(\epsilon)}_{\omega}$'s leads to the Hartle-Hawking-Israel state or the Minkowski vacuum in the flat spacetime. But it is clear that we need to know the $F^{(\epsilon)}_{\omega}$'s modes for doing the construction. With each mode there is associated an annihilation operator, the vacuum which is annihilated by the operators associated with the $H^{(\epsilon)}_{\omega}$'s is the Hartle-Hawking-Israel state. For more details see the paper of Israel \cite{is76}.

\end{document}